\documentclass[twocolumn,prl,aps,showpacs,preprintnumbers,amsmath,amssymb,twoside,floatfix]{revtex4-1}
\usepackage{graphicx}		% Include figure files
\usepackage{pstricks}
\usepackage{dcolumn}		% Align table columns on decimal point
\usepackage{bm}			% bold math
\usepackage{units}		% Formating number and unit
\usepackage{color}		% text color possible

\begin{document}

% used chemical formulars
\newcommand{\YMO}{YMn$_{\mbox{\scriptsize 2}}$O$_{\mbox{\scriptsize 5}}$}
\newcommand{\TMO}{TbMn$_{\mbox{\scriptsize 2}}$O$_{\mbox{\scriptsize 5}}$}
%
% resonant edges
\newcommand{\MnL}{Mn\,$L_{\mbox{\scriptsize 2,3}}$}
\newcommand{\MnLt}{Mn\,$L_{\mbox{\scriptsize 3}}$}
\newcommand{\OK}{O\,$K$}
%
% modulation vectors
\newcommand{\icp}{$(0.48, 0, 0.29)$}
\newcommand{\cp}{$(1/2, 0, 1/4)$}
%
% electric polarization
\newcommand{\Pfe}{$\mathcal{P}$}
\newcommand{\Pb}{$\mathcal{P}_b$}
\newcommand{\Pion}{$\mathcal{P}_{\mathrm{ion}}$}
\newcommand{\Pel}{$\mathcal{P}_{\mathrm{el}}$}
%
% Fermi energy
%\newcommand{\EF}{$\mathcal{E}_\mathrm{F}$}
\newcommand{\EF}{$\epsilon_\mathrm{F}$}
% x-ray / X-ray
\newcommand{\xray}{X-ray}
%
% states
\newcommand{\op}{O\,$2p$}
\newcommand{\mnd}{Mn\,$3d$}

\title{Observation of Electronic Ferroelectric Polarization in Multiferroic YMn$_2$O$_5$}

\author{S.~Partzsch$^1$}
\author{S.B.~Wilkins$^2$}
\author{J.P.~Hill$^2$}
\author{E.~Schierle$^3$}
\author{E.~Weschke$^3$}
\author{D.~Souptel$^1$}
\author{B.~B\"uchner$^{1}$}
\author{J.~Geck$^1$}

\affiliation{$^1$Leibniz Institute for Solid State and Materials Research IFW
Dresden, Helmholtzstrasse 20, 01069 Dresden, Germany}

\affiliation{$^2$Condensed Matter Physics and Material Science Department, Brookhaven National Laboratory, Upton, New York, 11973 USA}%

\affiliation{$^3$Helmholtz-Zentrum Berlin f\"ur Materialien und Energie,
Albert-Einstein-Str.\,15, 12489 Berlin, Germany}

\date{Received: \today}
% / Revised version: date}

\begin{abstract}
We report the observation of a  magnetic polarization of the O\,$2p$-states in YMn$_2$O$_5$ through the use of soft X-ray resonant scattering at the oxygen $K$-edge. Remarkably, we find that the temperature dependence of the integrated intensity of this signal closely follows the macroscopic electric polarization, and hence is proportional to the ferroelectric order parameter. This is in contrast to the temperature dependence observed at the Mn\,$L_3$-edge, which reflects the Mn magnetic order parameter. First principle calculations provide a microscopic understanding of these results and show that a spin-dependent hybridization of O\,$2p$- and Mn\,$3d$-states results in a purely electronic contribution to the ferroelectric polarization, which can exist in the absence of lattice distortions.
\end{abstract}

% 75.85.+t : Magnetoelectric effects, multiferroics
% 77.80.-e : Ferroelectricity and antiferroelectricity
% 64.70.Rh : Commensurate-incommensurate transitions
% 61.05.C- : X-ray diffraction and scattering

\maketitle

%\section{Introduction}
% multiferroics
Multiferroic materials with a coupled ordering of electric and magnetic moments are extremely interesting for technological applications. They could, for example, be used to build a magnetic computer memory that could be switched with an electrical field (see e.\,g. \cite{Cheong2007,Khomskii2009}).
This has motivated a long history of research on these materials. However, compounds with the desired properties are very rare. Magnetism and ferroelectricity seldom coexist and even when they do, the coupling between them is typically extremely weak. The discovery of a number of transition metal oxides (TMOs) with a strong coupling \cite{Kimura2003,Hur2004} has therefore generated much excitement.

Attention has quickly turned to understanding the mechanism for this coupling.
Here, models based on structural distortions, which optimize the energy of the magnetic system, play a central role \cite{Cheong2007,Khomskii2009,Kimura2003,Hur2004}. Since these scenarios focus on magnetically driven distortions of an ionic lattice, we will refer to
the corresponding ionic ferroelectric (FE) polarization as \Pion .

Transition metal oxides are, however, notorious for having strong electronic correlations and interactions that not only involve lattice and spins, but also charges and orbitals.
Further, it is well known that covalency between the transition metal ion and the oxygen plays a major role in most TMOs \cite{ZaanenSawatzkyAllen1985}.
This has led to the suggestion that there may be another contribution to the polarization, i.\,e., one due to the valence electrons.
Recent theoretical studies point to such an electronic contribution, \Pel\/ to the FE polarization \cite{Moskvin2008, Giovannetti2008}.
In contrast to \Pion , \Pel\/ is not primarily related to structural distortions, but rather to the covalency between the oxygen and the transition metal, i.\,e., it is a polarization that can exist even if there is no movement of the ion cores from their undistorted positions. Whether or not such a \Pel\/ is in fact significant in these materials is currently hotly debated. While there is some indirect evidence for it \cite{Lottermoser2009}, to date, direct experimental tests of this picture have been lacking.

\begin{figure}[b!]
\center{
\resizebox{0.85\columnwidth}{!}{%
 \includegraphics[clip,angle=0]{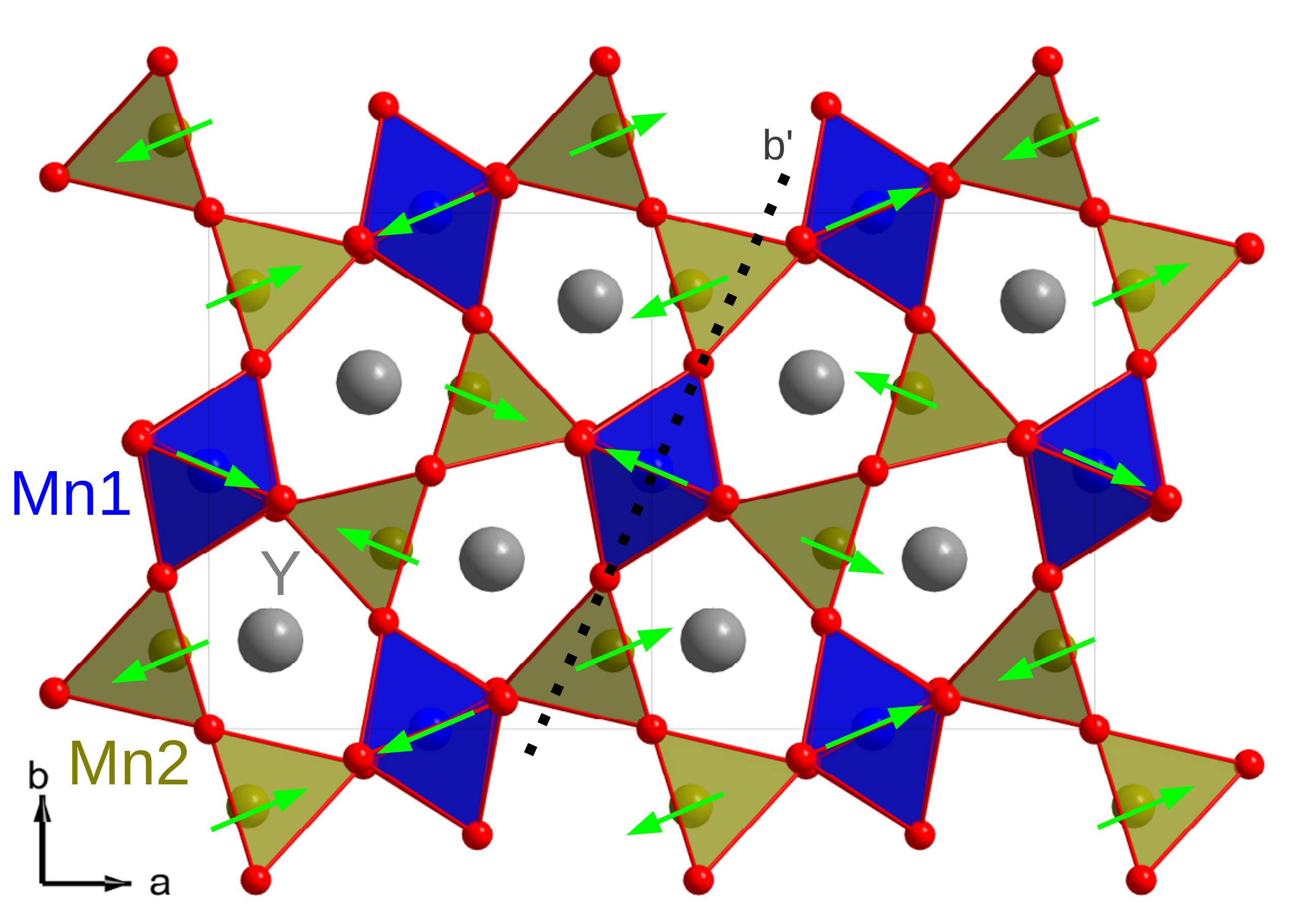}}
}
\caption{In \YMO\/ Mn1-octahedral chains (blue) in the $c$-direction are linked by Mn2-pyramidal dimers (brown) in the $ab$-plane. The grey spheres are Y, the red O \cite{Quezel-Ambrunaz1964,Tachibana2005}. The green arrows represent the commensurate spin structure at $\unit[25]{K}$ as determined by neutron diffraction \cite{Noda2006,Vecchini2008}.
The dashed line in the center indicates the plane shown in Fig.\,\ref{lsda-fig}.}
\label{struc-fig}
\end{figure}

% outline of paper
In this letter, we report the observation of \Pel, in multiferroic \YMO\/ by means of resonant soft \xray\/ scattering (RSXS) at the \OK-edge. The RSXS signal at the \OK-edge is due to a redistribution of valence charge, resulting in a magnetic polarization of the \op-states. % and the \Pel\/ in this material.
The corresponding magnetic signal at the \OK-edge does not follow the magnetic scattering from the Mn moments, but rather closely follows the ferroelectric polarization. Employing density functional theory (DFT) calculations, we provide a microscopic explanation of this observation and show that strongly spin-dependent hybridizations result in a purely electronic contribution \Pel\/ to the macroscopic FE polarization \Pfe. Our RSXS measurements at the \OK-edge are measurements of this \Pel . We emphasize that \Pel\/ is possible even in the absence of any lattice distortion.

% YMn2O5 crystal structure
The crystal structure of \YMO\ is illustrated in Fig.\,\ref{struc-fig}. Manganese occupies two different crystallographic sites; the first coordinated by an oxygen octahedron (Mn1) and the second by an oxygen square-based pyramid (Mn2) \cite{Quezel-Ambrunaz1964,Tachibana2005}. The Mn1-octahedra form edge-sharing chains along the $c$-direction that are linked within the $ab$-plane by Mn2-pyramid dimers. %(Fig.\/\ref{struc-fig}). 
Within an ionic description, the very different local coordination of Mn1 and Mn2 results in different formal valencies of these two sites, namely 4+ ($3d^3$) for Mn1 and 3+ ($3d^4$) for Mn2 \cite{Abrahams1967}.

% YMn2O5 physical properties
As a function of temperature, \YMO\/ undergoes a sequence of phase transitions, which demonstrate the strong coupling between the magnetic and FE order: Upon cooling, a transition from  paramagnetic to incommensurate antiferromagnetic (HTIC-phase) occurs at $T_1=\unit[45]{K}$ \cite{Kagomiya2001}. At $T_2=\unit[39]{K}$  the antiferromagnetic order becomes commensurate (C-phase) \cite{Kagomiya2001} and a finite \Pfe\/ appears \cite{Noda2003}.
Upon further cooling, low-temperature incommensurate magnetic order sets in at $T_3=\unit[19]{K}$ (LTIC-phase) \cite{Kagomiya2001}, which is connected to a sign change and a reduction in magnitude of  \Pfe\, \cite{Noda2003}.

\begin{figure}[t!]
\center{
\resizebox{0.9\columnwidth}{!}{%
 \includegraphics[clip,angle=0]{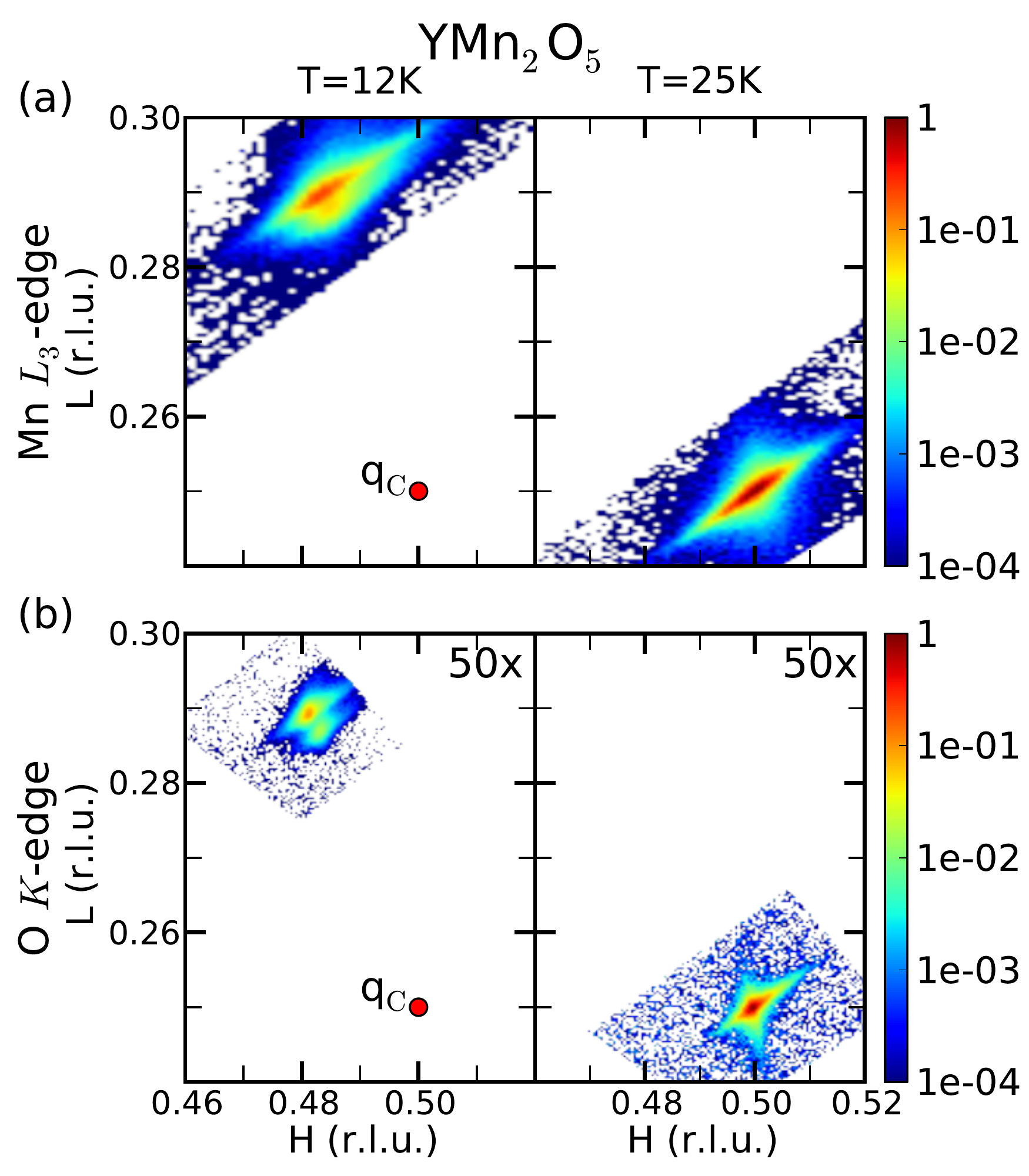}
}
}
\caption{(a) and (b) show the resonant \xray\/ scattering in the $(H0L)$-plane as measured at the \MnLt-edge ($\unit[644.6]{eV}$) and \OK-edge ($\unit[529.6]{eV}$), respectively.
The observed peak moves from the commensurate position \cp\/ ($\unit[25]{K}$, right) to \icp\/ ($\unit[12]{K}$, left), in the incommensurate phase.}
\label{hl-fig}
\end{figure}

% experimental
The \YMO\/ single crystals used for our study were grown using the PbO-PbF$_2$ flux method \cite{Wanklyn1972} and characterized by specific heat, magnetization and \xray\/ diffraction measurements, which were all in excellent agreement with previously published results \cite{Ikeda1995,Quezel-Ambrunaz1964,Tachibana2005}. For the RSXS experiments, a \YMO\/ sample with a (201)-surface normal was polished using $\unit[0.1]{\mu m}$ diamond films.
Experiments were performed at the UE46-PGM1 beamline of BESSY II at the Helmholtz-Zentrum Berlin and the X1A2 beamline at the NSLS, Brookhaven National Laboratory. The former setup realizes a horizontal scattering experiment with variable incoming polarization and is equipped with a photodiode point detector. The scattering plane of the NSLS-chamber was vertical and the incoming polarization horizontal. In this case, two-dimensional cuts in reciprocal space were recorded using a Princeton Instruments PI-MTE CCD. In each case, the sample was cooled using a liquid helium cryostat.

% hl-plane HTIC/C MnL3
In Fig.\,\ref{hl-fig}, RSXS-intensity distributions in the $(H0L)$-plane are presented for the LTIC- (left column) and the C-phase (right column). The two $(H0L)$-maps in the upper row of the figure were taken  with the photon energy tuned to the main absorption peak of the \MnLt-edge at $\unit[644.6]{eV}$ \cite{Staub2010}. At this photon energy RSXS becomes extremely sensitive to magnetic order on the Mn-sites \cite{Wilkins-327-SRXS-2003}.
In the C-phase, we observed a wavevector of \cp\/ which was in agreement with the magnetic wave-vector determined by neutron diffraction \cite{Kagomiya2001}. In the LTIC phase, an incommensurate wavevector \icp \/ was observed, again in very good agreement with previous neutron diffraction measurements \cite{Kagomiya2001,Radaelli2009}.

% hl-plane HTIC/C OK
More surprisingly, and more importantly, we also observed a superlattice reflection at the magnetic wave-vector when the photon energy was tuned to the \OK-edge at $\unit[529.6]{eV}$ \cite{Beale2010}. This is shown in Fig.\,\ref{hl-fig}\,(b). The transition from the C-phase into the LTIC-phase is also clearly seen at the \OK-edge, as demonstrated in Fig.\,\ref{hl-fig}\,(b). We attribute the scattering at the \OK-edge to a magnetic polarization of the \op\/ states. This is for two reasons. First, the wavevectors measured at the \OK-edge and the \MnL-edge are identical. Second, a magnetic polarization of the \op\/ states was also found in previous measurements on the iso-structural compound \TMO\/ \cite{Beale2010}.
\xray\ magnetic circular dichroism measurements (XMCD) on related transition metal oxides have shown that the magnetic signal at the oxygen K-edge can be interpreted as an orbital moment transfered to the oxygen via strong hybridization between the \mnd\/ and \op\/ states \cite{Pellegrin1997, Goering2002}, consistent with our results.

% peak widths and penetration depths
The width of the superlattice peak along $\mathbf{q}$ is about three times larger at the \MnLt-edge than at the \OK-edge. This is simply due to the different \xray\/ penetration depths at the two edges. An estimate from our XAS spectra and the tabulated values of Henke \cite{Henke1993}, yields \xray\/ penetration depths of $\unit[0.04]{\mu m}$ and $\unit[0.12]{\mu m}$ at the \MnLt- and \OK-edges, respectively, entirely consistent with the different observed peak widths.

\begin{figure}[t!]
\center{
\resizebox{0.85\columnwidth}{!}{%
 \includegraphics[clip,angle=0]{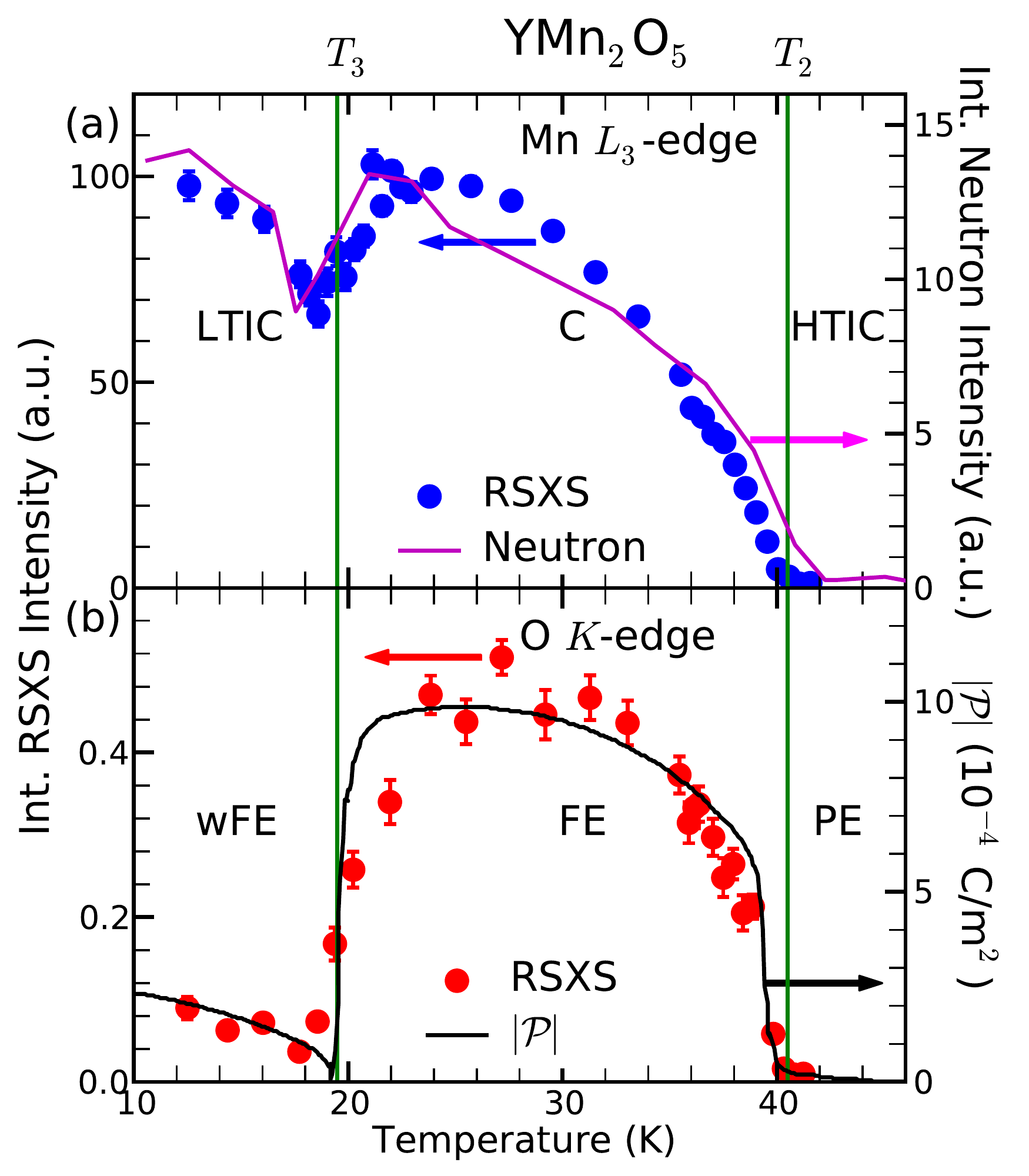}
}
}
\caption{Temperature dependence of the soft \xray\/ integrated intensity as measured at the \MnLt- (a) and \OK-edges (b).
Good agreement is seen with the reported (solid lines)
integrated intensity observed by neutron diffraction data (a) \cite{Kagomiya2001} and
magnitude of the electric polarization $|$\Pfe$|$ (b) \cite{Noda2003}, respectively.
The phases are low temperature incommensurate (LTIC) weak ferroelectric (wFE), commensurate (C) ferroelectric (FE) and high temperature incommensurate (HTIC) paraelectric (PE).}
\label{it-fig}
\end{figure}

% I(T) Mn L3-edge
We now turn to the temperature dependences of the superlattice reflections as measured at the \MnLt-edge and the \OK-edge. These are shown in Fig.\,\ref{it-fig}. In each case, the intensity of the reflections was accurately and efficiently integrated over a full 3D volume in reciprocal space using the CCD camera.
% I(T) Mn L3-edge
The integrated intensity measured at the \MnLt-edge is compared to the corresponding magnetic neutron scattering data \cite{Kagomiya2001} in Fig.\,\ref{it-fig}\,(a).
The very good agreement between the two confirms that the RSXS
signal at the \MnLt-edge is measuring the square of the
magnetic order parameter - just as neutron scattering does.

% I(T) O K-edge
As can be seen in Fig.\,\ref{it-fig}\,(b), the temperature dependence observed at the \OK-edge is quite different and does not follow the magnetic order parameter. Instead, it drops dramatically at the C-LTIC transition and does not recover at low temperatures, reaching only about 20\% of its maximum value in the commensurate phase. This is the central result of this study:  the RSXS signal at the \OK-edge closely follows the absolute value of the macroscopic ferroelectric polarization  $|\mathcal{P}|$, which is plotted in Fig.\,\ref{it-fig}\,(b) \cite{Noda2003}. Within the error of our experiment, the two quantities have identical temperature dependences, even through the transition from the C-phase into the LTIC-phase. This establishes a direct relation between the electronic modulation of the \op\/ states and the ferroelectricity of \YMO.

% LSDA+U method
Motivated by these results, we have performed LSDA+U calculations in order to clarify this relationship. The {\it ab initio} calculations were performed employing the full potential {\tt WIEN2K} code \cite{Wien2K}, using 250 k-points in the irreducible Brillouin zone. We set $U=\unit[5]{eV}$ and $J=\unit[0.88]{eV}$ (cf. Ref.\,\onlinecite{Giovannetti2008}) and performed calculations without spin-orbit interactions.
%
% no observed structural changes in refinements
The calculations were based on the atomic positions, which we determined by means of standard single crystal \xray\/ diffraction for our samples in the C-phase.
The resulting structural refinements are in very good agreement with previous studies \cite{Quezel-Ambrunaz1964,Tachibana2005} and gave $Pbam$ as the space group of \YMO. As in previous studies, no additional structural distortion or symmetry reduction could be resolved in the C-phase \cite{Noda2007}.
%
% subgroup
Therefore we chose the maximal non-isomorphic subgroup of $Pbam$ that allows for the experimentally observed \Pfe \/ along the $b$-direction, namely $Pb2_1m$ \cite{Kagomiya2003}.

% approximated magnetic structure
Since the magnetic unit cell of the real material is large and we are only interested in qualitative effects at this point, we studied a simplified spin structure, which approximates the observed C-phase and can still be described using the $Pb2_1m$ unit cell. In this symmetry, all the Mn1-sites are equivalent and were set to have parallel spins. In addition, there are two inequivalent Mn2-sites, Mn2a and Mn2b, in $Pb2_1m$. We set the spin of Mn2a to be parallel and that of Mn2b to be antiparallel to the Mn1-spin. 

\begin{figure}[b!]
\center{
%\showthe\columnwidth
\resizebox{0.8\columnwidth}{!}{%
 \includegraphics[clip,angle=0]{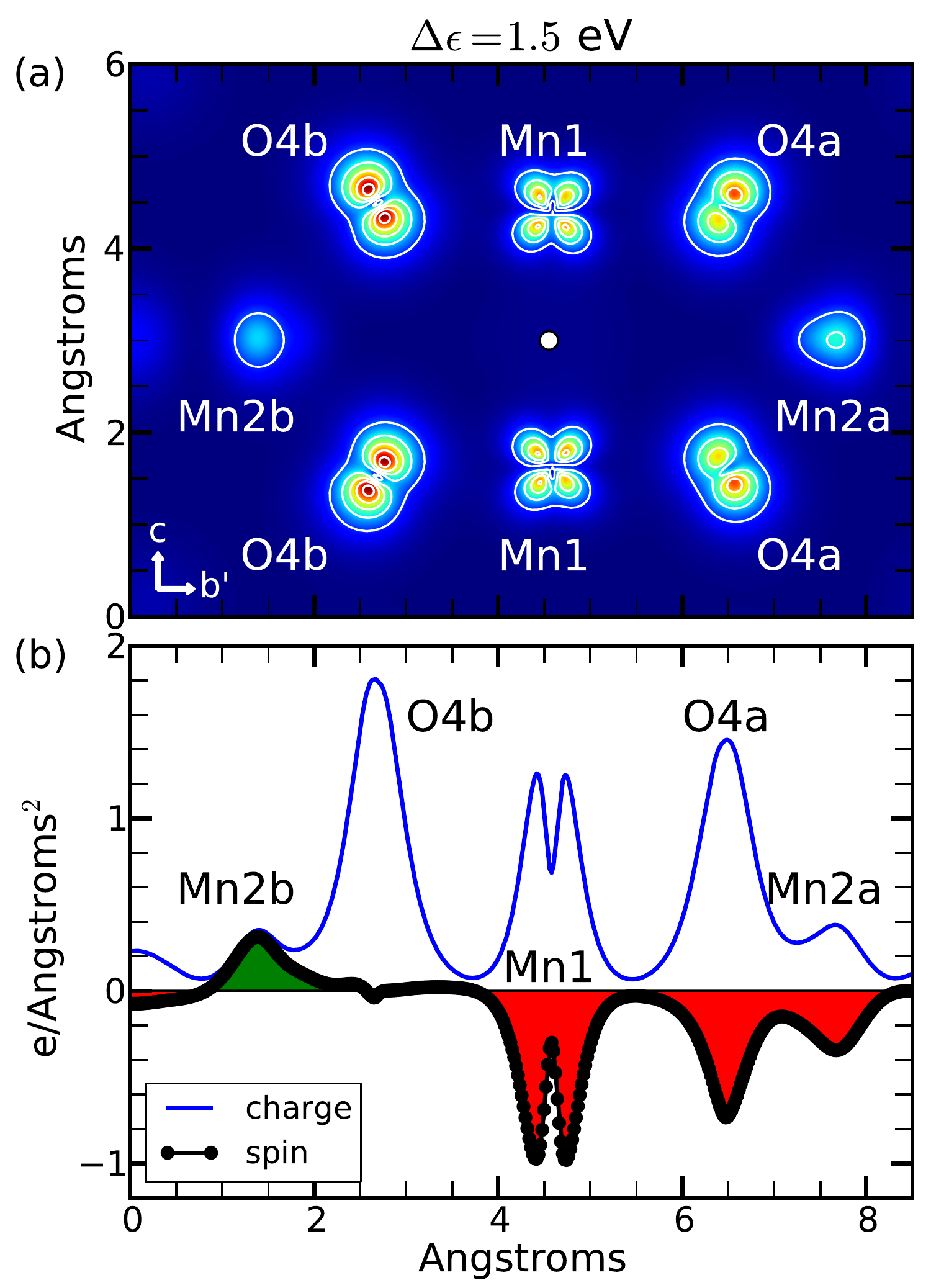}
}
}
\caption{Results from LSDA+U calculations in $Pb2_1m$ symmetry.
(a) Charge density $\rho_c=\rho(\uparrow)+\rho(\downarrow)$ of the valence states within 1.5\,eV of the Fermi level in the $b'c$-plane. The $b'$-axis is tilted by 23.5$^\circ$ from the $b$-axis (Fig.\,\ref{struc-fig}). (b) Projections of $\rho_c$ shown in (a) (blue line) and the corresponding spin density $\rho_s=\rho(\uparrow)-\rho(\downarrow)$ (filled curve) onto the $b'$-axis.}
\label{lsda-fig}
\end{figure}

% Discussion of b-chain, basal Mn1-O4-Mn2 path
We focus on the Mn2a--Mn1--Mn2b bond, involving only oxygens in the basal plane of the Mn2-pyramids. As illustrated in Fig.\,\ref{struc-fig}, the corresponding spins in the C-phase of the real material have $\uparrow\downarrow\downarrow$ projections onto the $a$-direction. This is also the situation described by our model calculation.
The LSDA+U charge density and net spin density within the plane of the Mn2a--Mn1--Mn2b bond are illustrated in Figs.\,\ref{lsda-fig}\,(a) and (b). The densities in these figures correspond to the valence electrons within 1.5\,eV of the Fermi energy.
%
% O K-edge RSXS <-> magnetic polarization <-> hybridization <-> Pel
As can clearly be observed in these figures, the $\uparrow\downarrow\downarrow$ spin configuration corresponds to a spatial distribution of the valence charge density, which breaks the center of inversion [circle in Fig.\,\ref{lsda-fig}\,(a)]. In particular, the hybridization within a Mn2--O--Mn1 bond depends on the relative Mn-spin orientation.
This spin-dependent hybridization has a large effect on the oxygen states and yields a ferroelectric moment along $b$, as observed experimentally.
The RSXS signal at the \OK-edge, which measures the magnetic polarization of oxygen, is directly related to this spin-dependent Mn2--O--Mn1 hybridization and, hence, reflects \Pel.

% no Pion in first calculation
The above calculations do not take into account spin orbit interactions or structural distortions. Thus the FE polarization in this model results purely from the spin-dependent Mn2--O--Mn1 hybridization and is due entirely to a redistribution of valence charge. The atomic positions are not relaxed and therefore, by definition, there is no \Pion. The important result therefore is that, within LSDA+U, a purely electronic contribution to the FE polarization exists, which involves a spatial modulation of the  \op\/ valence states (Fig.\,\ref{lsda-fig}).

% influence of structural distortions
We also checked the influence of structural distortions within LSDA+U and found that they also affect the Mn-O covalency. However, in our calculations this effect is smaller than that due to the spin-induced valence charge redistribution, because the changes in atomic positions are very small. They are estimated to be of the order of \unit[0.01]{\AA} or less\,\,\cite{Giovannetti2008,Noda2003} and, in fact, have not be resolved experimentally to date \cite{Noda2007}.

% comparison with Giovannetti2008
The LSDA+U calculations presented here agree very well with a more detailed DFT study of the related material HoMn$_2$O$_5$\,\cite{Giovannetti2008}, which also revealed a covalency-driven \Pel\/ and showed that this contribution partially cancels the ionic \Pion\/.
The present experiments are confirmation of the theoretical prediction and indicate that \Pel\/ is a general and important feature of these compounds.

% P = Pel + Pion T-dep
We remark that the fact that \Pel\/, as measured in our experiments,  and \Pfe\/ have the same temperature dependence (Fig.\,2) indicates that either \Pel\/ is the dominant contribution to \Pfe, or that \Pel\/ and \Pion\/ have the same temperature dependence since \Pfe = \Pion + \Pel. From our experiments we cannot \emph{a priori} determine which of these is true.

%\section{Summary}
In summary, we have performed RSXS experiments on \YMO, revealing a spatial modulation of the \op\/ states, which is tightly coupled to the magnetic order, displaying the same periodicity and undergoing changes at the magnetic transition. Most importantly, the integrated intensity
of the oxygen superstructure reflection is directly proportional to the macroscopic FE polarization.
The present experiments verify that \Pel\/ does indeed exist in a real material and imply that Mn-O covalency and the resulting \Pel\/ play a major role for the ferroelectricity in \YMO.

% acknowledgement
We thank J.-E. Hamann-Borrero and H. Wadati for their help with the experiments at UE46, and D. S. Coburn, W. Leonhardt, W. Schoenig and S. Wirick for technical support at X1A2. S.P. and J.G. thank the DFG for the support through the Emmy Noether Program GE 1647/2-1. Work performed at BNL was supported by the US Department of Energy, Division of Materials Science, under contract No. DE-AC02-98CH10886.

%bibtex
%\bibliographystyle{apsrev4-1}
%\bibliography{References/RMn2O5,References/Multiferroics,References/RMnO3,References/diffraction,References/electronic_structure}

\begin{thebibliography}{27}%
\makeatletter
\providecommand \@ifxundefined [1]{%
 \@ifx{#1\undefined}
}%
\providecommand \@ifnum [1]{%
 \ifnum #1\expandafter \@firstoftwo
 \else \expandafter \@secondoftwo
 \fi
}%
\providecommand \@ifx [1]{%
 \ifx #1\expandafter \@firstoftwo
 \else \expandafter \@secondoftwo
 \fi
}%
\providecommand \natexlab [1]{#1}%
\providecommand \enquote  [1]{``#1''}%
\providecommand \bibnamefont  [1]{#1}%
\providecommand \bibfnamefont [1]{#1}%
\providecommand \citenamefont [1]{#1}%
\providecommand \href@noop [0]{\@secondoftwo}%
\providecommand \href [0]{\begingroup \@sanitize@url \@href}%
\providecommand \@href[1]{\@@startlink{#1}\@@href}%
\providecommand \@@href[1]{\endgroup#1\@@endlink}%
\providecommand \@sanitize@url [0]{\catcode `\\12\catcode `\$12\catcode
  `\&12\catcode `\#12\catcode `\^12\catcode `\_12\catcode `\%12\relax}%
\providecommand \@@startlink[1]{}%
\providecommand \@@endlink[0]{}%
\providecommand \url  [0]{\begingroup\@sanitize@url \@url }%
\providecommand \@url [1]{\endgroup\@href {#1}{\urlprefix }}%
\providecommand \urlprefix  [0]{URL }%
\providecommand \Eprint [0]{\href }%
\providecommand \doibase [0]{http://dx.doi.org/}%
\providecommand \selectlanguage [0]{\@gobble}%
\providecommand \bibinfo  [0]{\@secondoftwo}%
\providecommand \bibfield  [0]{\@secondoftwo}%
\providecommand \translation [1]{[#1]}%
\providecommand \BibitemOpen [0]{}%
\providecommand \bibitemStop [0]{}%
\providecommand \bibitemNoStop [0]{.\EOS\space}%
\providecommand \EOS [0]{\spacefactor3000\relax}%
\providecommand \BibitemShut  [1]{\csname bibitem#1\endcsname}%
\let\auto@bib@innerbib\@empty
%</preamble>
\bibitem [{\citenamefont {Cheong}\ and\ \citenamefont
  {Mostovoy}(2007)}]{Cheong2007}%
  \BibitemOpen
  \bibfield  {author} {\bibinfo {author} {\bibfnamefont {S.-W.}\ \bibnamefont
  {Cheong}}\ and\ \bibinfo {author} {\bibfnamefont {M.}~\bibnamefont
  {Mostovoy}},\ }\href {http://dx.doi.org/10.1038/nmat1804} {\bibfield
  {journal} {\bibinfo  {journal} {Nature}\ }\textbf {\bibinfo {volume}
  {6}},\ \bibinfo {pages} {13} (\bibinfo {year} {2007})}\BibitemShut {NoStop}%
\bibitem [{\citenamefont {Khomskii}(2009)}]{Khomskii2009}%
  \BibitemOpen
  \bibfield  {author} {\bibinfo {author} {\bibfnamefont {D.}~\bibnamefont
  {Khomskii}},\ }\href {\doibase 10.1103/Physics.2.20} {\bibfield  {journal}
  {\bibinfo  {journal} {Physics}\ }\textbf {\bibinfo {volume} {2}},\ \bibinfo
  {eid} {20} (\bibinfo {year} {2009})}\BibitemShut {NoStop}%
\bibitem [{\citenamefont {Kimura}\ \emph {et~al.}(2003)\citenamefont {Kimura},
  \citenamefont {Ishihara}, \citenamefont {Shintani}, \citenamefont {Arima},
  \citenamefont {Takahashi}, \citenamefont {Ishizaka},\ and\ \citenamefont
  {Tokura}}]{Kimura2003}%
  \BibitemOpen
  \bibfield  {author} {\bibinfo {author} {\bibfnamefont {T.}~\bibnamefont
  {Kimura}}\ \emph {et~al.},\ }\href {\doibase
  10.1103/PhysRevB.68.060403} {\bibfield  {journal} {\bibinfo  {journal} {Phys.
  Rev. B}\ }\textbf {\bibinfo {volume} {68}},\ \bibinfo {pages} {060403}
  (\bibinfo {year} {2003})}\BibitemShut {NoStop}%
\bibitem [{\citenamefont {Hur}\ \emph {et~al.}(2004)\citenamefont {Hur},
  \citenamefont {Park}, \citenamefont {Sharma}, \citenamefont {Ahn},
  \citenamefont {Guha},\ and\ \citenamefont {Cheong}}]{Hur2004}%
  \BibitemOpen
  \bibfield  {author} {\bibinfo {author} {\bibfnamefont {N.}~\bibnamefont
  {Hur}}\ \emph {et~al.},\ }\href {\doibase 10.1038/nature02572} {\bibfield
  {journal} {\bibinfo  {journal} {Nature}\ }\textbf {\bibinfo {volume} {429}},\
  \bibinfo {pages} {392} (\bibinfo {year} {2004})}\BibitemShut {NoStop}%
\bibitem [{\citenamefont {Zaanen}\ \emph {et~al.}(1985)\citenamefont {Zaanen},
  \citenamefont {Sawatzky},\ and\ \citenamefont
  {Allen}}]{ZaanenSawatzkyAllen1985}%
  \BibitemOpen
  \bibfield  {author} {\bibinfo {author} {\bibfnamefont {J.}~\bibnamefont
  {Zaanen}}\ \emph {et~al.},\ }\href {\doibase 10.1103/PhysRevLett.55.418} {\bibfield  {journal}
  {\bibinfo  {journal} {Phys. Rev. Lett.}\ }\textbf {\bibinfo {volume} {55}},\
  \bibinfo {pages} {418} (\bibinfo {year} {1985})}\BibitemShut {NoStop}%
\bibitem [{\citenamefont {Moskvin}\ and\ \citenamefont
  {Pisarev}(2008)}]{Moskvin2008}%
  \BibitemOpen
  \bibfield  {author} {\bibinfo {author} {\bibfnamefont {A.~S.}\ \bibnamefont
  {Moskvin}}\ and\ \bibinfo {author} {\bibfnamefont {R.~V.}\ \bibnamefont
  {Pisarev}},\ }\href {\doibase 10.1103/PhysRevB.77.060102} {\bibfield
  {journal} {\bibinfo  {journal} {Phys. Rev. B}\ }\textbf {\bibinfo {volume}
  {77}},\ \bibinfo {pages} {060102} (\bibinfo {year} {2008})}\BibitemShut
  {NoStop}%
\bibitem [{\citenamefont {Giovannetti}\ and\ \citenamefont {van~den
  Brink}(2008)}]{Giovannetti2008}%
  \BibitemOpen
  \bibfield  {author} {\bibinfo {author} {\bibfnamefont {G.}~\bibnamefont
  {Giovannetti}}\ and\ \bibinfo {author} {\bibfnamefont {J.}~\bibnamefont
  {van~den Brink}},\ }\href {\doibase 10.1103/PhysRevLett.100.227603}
  {\bibfield  {journal} {\bibinfo  {journal} {Phys. Rev. Lett.}\
  }\textbf {\bibinfo {volume} {100}},\ \bibinfo {eid} {227603} (\bibinfo {year}
  {2008})}\BibitemShut {NoStop}%
\bibitem [{\citenamefont {Lottermoser}\ \emph {et~al.}(2009)\citenamefont
  {Lottermoser}, \citenamefont {Meier}, \citenamefont {Pisarev},\ and\
  \citenamefont {Fiebig}}]{Lottermoser2009}%
  \BibitemOpen
  \bibfield  {author} {\bibinfo {author} {\bibfnamefont {T.}~\bibnamefont
  {Lottermoser}}\ \emph {et~al.},\ }\href
  {\doibase 10.1103/PhysRevB.80.100101} {\bibfield  {journal} {\bibinfo
  {journal} {Phys. Rev. B}\
  }\textbf {\bibinfo {volume} {80}},\ \bibinfo {eid} {100101} (\bibinfo {year}
  {2009})}\BibitemShut {NoStop}%
\bibitem [{\citenamefont {Quezel-Ambrunaz}\ \emph {et~al.}(1964)\citenamefont
  {Quezel-Ambrunaz}, \citenamefont {Bertaut},\ and\ \citenamefont
  {Buison}}]{Quezel-Ambrunaz1964}%
  \BibitemOpen
  \bibfield  {author} {\bibinfo {author} {\bibfnamefont {S.}~\bibnamefont
  {Quezel-Ambrunaz}}\ \emph {et~al.},\ } {\bibfield  {journal} {\bibinfo  {journal} {Comp. Ren. Aca. sci.}\ }\textbf {\bibinfo {volume}
  {258}},\ \bibinfo {pages} {3025} (\bibinfo {year} {1964})}\BibitemShut
  {NoStop}%
\bibitem [{\citenamefont {Tachibana}\ \emph {et~al.}(2005)\citenamefont
  {Tachibana}, \citenamefont {Akiyama}, \citenamefont {Kawaji},\ and\
  \citenamefont {Atake}}]{Tachibana2005}%
  \BibitemOpen
  \bibfield  {author} {\bibinfo {author} {\bibfnamefont {M.}~\bibnamefont
  {Tachibana}}\ \emph {et~al.},\ }\href {\doibase
  10.1103/PhysRevB.72.224425} {\bibfield  {journal} {\bibinfo  {journal} {Phys.
  Rev. B}\ }\textbf {\bibinfo {volume} {72}},\ \bibinfo {pages} {224425}
  (\bibinfo {year} {2005})}\BibitemShut {NoStop}%
\bibitem [{\citenamefont {Noda}\ \emph {et~al.}(2006)\citenamefont {Noda},
  \citenamefont {Kimura}, \citenamefont {Kamada}, \citenamefont {Osawa},
  \citenamefont {Fukuda}, \citenamefont {Ishikawa}, \citenamefont {Kobayashi},
  \citenamefont {Wakabayashi}, \citenamefont {Sawa}, \citenamefont {Ikeda},\
  and\ \citenamefont {Kohn}}]{Noda2006}%
  \BibitemOpen
  \bibfield  {author} {\bibinfo {author} {\bibfnamefont {Y.}~\bibnamefont
  {Noda}}\ \emph {et~al.},\
  }\href {\doibase DOI: 10.1016/j.physb.2006.05.179} {\bibfield  {journal}
  {\bibinfo  {journal} {Phys. B: Cond. Mat.}\ }\textbf {\bibinfo
  {volume} {385-386}},\ \bibinfo {pages} {119 } (\bibinfo {year} {2006})}
  \BibitemShut {NoStop}%
\bibitem [{\citenamefont {Vecchini}\ \emph {et~al.}(2008)\citenamefont
  {Vecchini}, \citenamefont {Chapon}, \citenamefont {Brown}, \citenamefont
  {Chatterji}, \citenamefont {Park}, \citenamefont {Cheong},\ and\
  \citenamefont {Radaelli}}]{Vecchini2008}%
  \BibitemOpen
  \bibfield  {author} {\bibinfo {author} {\bibfnamefont {C.}~\bibnamefont
  {Vecchini}}\ \emph {et~al.},\ }\href {\doibase
  10.1103/PhysRevB.77.134434} {\bibfield  {journal} {\bibinfo  {journal}
  {Phys. Rev. B}\ }\textbf
  {\bibinfo {volume} {77}},\ \bibinfo {eid} {134434} (\bibinfo {year}
  {2008})}\BibitemShut {NoStop}%
\bibitem [{\citenamefont {Abrahams}\ and\ \citenamefont
  {Bernstein}(1967)}]{Abrahams1967}%
  \BibitemOpen
  \bibfield  {author} {\bibinfo {author} {\bibfnamefont {S.~C.}\ \bibnamefont
  {Abrahams}}\ and\ \bibinfo {author} {\bibfnamefont {J.~L.}\ \bibnamefont
  {Bernstein}},\ }\href {\doibase 10.1063/1.1840450} {\bibfield  {journal}
  {\bibinfo  {journal} {J. Chem. Phys.}\ }\textbf {\bibinfo
  {volume} {46}},\ \bibinfo {pages} {3776} (\bibinfo {year}
  {1967})}\BibitemShut {NoStop}%
\bibitem [{\citenamefont {Kagomiya}\ \emph {et~al.}(2001)\citenamefont
  {Kagomiya}, \citenamefont {Kimura}, \citenamefont {Noda},\ and\ \citenamefont
  {Kohn}}]{Kagomiya2001}%
  \BibitemOpen
  \bibfield  {author} {\bibinfo {author} {\bibfnamefont {I.}~\bibnamefont
  {Kagomiya}}\ \emph {et~al.},\ }\href
  {http://wwwsoc.nii.ac.jp/jps/jpsj/pdf/70A/70A-145.pdf} {\bibfield  {journal}
  {\bibinfo  {journal} {J. Phys. Soc. Jap.}\
  }\textbf {\bibinfo {volume} {70A}},\ \bibinfo {pages} {145} (\bibinfo {year}
  {2001})}\BibitemShut {NoStop}%
\bibitem [{\citenamefont {Noda}\ \emph {et~al.}(2003)\citenamefont {Noda},
  \citenamefont {Fukuda}, \citenamefont {Kimura}, \citenamefont {Kagomiya},
  \citenamefont {Matumoto}, \citenamefont {Kohn}, \citenamefont {Shobu},\ and\
  \citenamefont {Ikeda}}]{Noda2003}%
  \BibitemOpen
  \bibfield  {author} {\bibinfo {author} {\bibfnamefont {Y.}~\bibnamefont
  {Noda}}\ \emph {et~al.},\ }\href
  {http://www.kps.or.kr/home/kor/journal/library/search_list.asp} {\bibfield
  {journal} {\bibinfo  {journal} {J. Kor. Phys. Soc.}\
  }\textbf {\bibinfo {volume} {42}},\ \bibinfo {pages} {1192} (\bibinfo {year}
  {2003})}\BibitemShut {NoStop}%
\bibitem [{\citenamefont {Wanklyn}(1972)}]{Wanklyn1972}%
  \BibitemOpen
  \bibfield  {author} {\bibinfo {author} {\bibfnamefont {B.~M.}\ \bibnamefont
  {Wanklyn}},\ }\href {\doibase 10.1007/BF00549910} {\bibfield  {journal}
  {\bibinfo  {journal} {J. Mat. Sci.}\ }\textbf {\bibinfo
  {volume} {7}},\ \bibinfo {pages} {813} (\bibinfo {year} {1972})}\BibitemShut
  {NoStop}%
\bibitem [{\citenamefont {Ikeda}\ and\ \citenamefont {Kohn}(1995)}]{Ikeda1995}%
  \BibitemOpen
  \bibfield  {author} {\bibinfo {author} {\bibfnamefont {A.}~\bibnamefont
  {Ikeda}}\ and\ \bibinfo {author} {\bibfnamefont {K.}~\bibnamefont {Kohn}},\
  }\href {\doibase 10.1080/00150199508217317} {\bibfield  {journal} {\bibinfo
  {journal} {Ferroel.}\ }\textbf {\bibinfo {volume} {169}},\ \bibinfo
  {pages} {75} (\bibinfo {year} {1995})}\BibitemShut {NoStop}%
\bibitem [{\citenamefont {Staub}\ \emph {et~al.}(2010)\citenamefont {Staub},
  \citenamefont {Bodenthin}, \citenamefont {Garc\'\i{}a-Fern\'andez},
  \citenamefont {de~Souza}, \citenamefont {Garganourakis}, \citenamefont
  {Golovenchits}, \citenamefont {Sanina},\ and\ \citenamefont
  {Lushnikov}}]{Staub2010}%
  \BibitemOpen
  \bibfield  {author} {\bibinfo {author} {\bibfnamefont {U.}~\bibnamefont
  {Staub}}\ \emph {et~al.},\ }\href {\doibase 10.1103/PhysRevB.81.144401} {\bibfield
  {journal} {\bibinfo  {journal} {Phys. Rev. B}\ }\textbf {\bibinfo {volume}
  {81}},\ \bibinfo {pages} {144401} (\bibinfo {year} {2010})}\BibitemShut
  {NoStop}%
\bibitem [{\citenamefont {Wilkins}\ \emph {et~al.}(2003)\citenamefont
  {Wilkins}, \citenamefont {Hatton}, \citenamefont {Roper}, \citenamefont
  {Prabhakaran},\ and\ \citenamefont {Boothroyd}}]{Wilkins-327-SRXS-2003}%
  \BibitemOpen
  \bibfield  {author} {\bibinfo {author} {\bibfnamefont {S.~B.}\ \bibnamefont
  {Wilkins}}\ \emph {et~al.},\ }\href {\doibase
  10.1103/PhysRevLett.90.187201} {\bibfield  {journal} {\bibinfo  {journal}
  {Phys. Rev. Lett.}\ }\textbf {\bibinfo {volume} {90}},\ \bibinfo {pages}
  {187201} (\bibinfo {year} {2003})}\BibitemShut {NoStop}%
\bibitem [{\citenamefont {Radaelli}\ \emph {et~al.}(2009)\citenamefont
  {Radaelli}, \citenamefont {Vecchini}, \citenamefont {Chapon}, \citenamefont
  {Brown}, \citenamefont {Park},\ and\ \citenamefont {Cheong}}]{Radaelli2009}%
  \BibitemOpen
  \bibfield  {author} {\bibinfo {author} {\bibfnamefont {P.~G.}\ \bibnamefont
  {Radaelli}}\ \emph {et~al.},\ }\href {\doibase
  10.1103/PhysRevB.79.020404} {\bibfield  {journal} {\bibinfo  {journal}
  {Phys. Rev. B}\ }\textbf
  {\bibinfo {volume} {79}},\ \bibinfo {eid} {020404} (\bibinfo {year}
  {2009})}\BibitemShut {NoStop}%
\bibitem [{\citenamefont {Beale}\ \emph {et~al.}(2010)\citenamefont {Beale},
  \citenamefont {Wilkins}, \citenamefont {Johnson}, \citenamefont {Bland},
  \citenamefont {Joly}, \citenamefont {Forrest}, \citenamefont {McMorrow},
  \citenamefont {Yakhou}, \citenamefont {Prabhakaran}, \citenamefont
  {Boothroyd},\ and\ \citenamefont {Hatton}}]{Beale2010}%
  \BibitemOpen
  \bibfield  {author} {\bibinfo {author} {\bibfnamefont {T.}~\bibnamefont
  {Beale}}\ \emph {et~al.},\ }\href
  {http://prl.aps.org/abstract/PRL/v105/i8/e087203} {\bibfield  {journal}
  {\bibinfo  {journal} {Phys. Rev. Lett.}\ }\textbf {\bibinfo {volume} {105}},\
  \bibinfo {pages} {087203} (\bibinfo {year} {2010})}\BibitemShut {NoStop}%
\bibitem [{\citenamefont {Pellegrin}\ \emph {et~al.}(1997)\citenamefont
  {Pellegrin}, \citenamefont {Tjeng}, \citenamefont {de~Groot}, \citenamefont
  {Hesper}, \citenamefont {Sawatzky}, \citenamefont {Moritomo},\ and\
  \citenamefont {Tokura}}]{Pellegrin1997}%
  \BibitemOpen
  \bibfield  {author} {\bibinfo {author} {\bibfnamefont {E.}~\bibnamefont
  {Pellegrin}}\ \emph {et~al.},\ }\href {\doibase
  10.1051/jp4/1997030} {\bibfield  {journal} {\bibinfo  {journal} {J. Phys. IV
  France}\ }\textbf {\bibinfo {volume} {7}},\ \bibinfo {pages} {C2} (\bibinfo
  {year} {1997})}\BibitemShut {NoStop}%
\bibitem [{\citenamefont {Goering}\ \emph {et~al.}(2002)\citenamefont
  {Goering}, \citenamefont {Bayer}, \citenamefont {Gold}, \citenamefont
  {Sch\"utz}, \citenamefont {Rabe}, \citenamefont {R\"udiger},\ and\
  \citenamefont {G\"untherodt}}]{Goering2002}%
  \BibitemOpen
  \bibfield  {author} {\bibinfo {author} {\bibfnamefont {E.}~\bibnamefont
  {Goering}}\ \emph {et~al.},\ }\href
  {http://stacks.iop.org/0295-5075/58/i=6/a=906} {\bibfield  {journal}
  {\bibinfo  {journal} {Europhys. Lett.}\ }\textbf {\bibinfo {volume}
  {58}},\ \bibinfo {pages} {906} (\bibinfo {year} {2002})}\BibitemShut
  {NoStop}%
\bibitem [{\citenamefont {Henke}\ \emph {et~al.}(1993)\citenamefont {Henke},
  \citenamefont {Gullikson},\ and\ \citenamefont {Davis}}]{Henke1993}%
  \BibitemOpen
  \bibfield  {author} {\bibinfo {author} {\bibfnamefont {B.~L.}\ \bibnamefont
  {Henke}}\ \emph {et~al.},\ }\href {\doibase DOI: 10.1006/adnd.1993.1013} {\bibfield
  {journal} {\bibinfo  {journal} {Atom. Data Nuc. Data Tab.}\
  }\textbf {\bibinfo {volume} {54}},\ \bibinfo {pages} {181 } (\bibinfo {year}
  {1993})}\BibitemShut {NoStop}%
\bibitem [{\citenamefont {Blaha}\ \emph {et~al.}(2001)\citenamefont {Blaha},
  \citenamefont {Schwarz}, \citenamefont {Madsen}, \citenamefont {Kvasnicka},\
  and\ \citenamefont {Luitz}}]{Wien2K}%
  \BibitemOpen
  \bibfield  {author} {\bibinfo {author} {\bibfnamefont {P.}~\bibnamefont
  {Blaha}}\ \emph {et~al.},\ } {\emph
  {\bibinfo {title} {An Augmented Plane Wave + Local Orbitals Program for
  Calculating Crystal Properties}}}\ (\bibinfo  {publisher} {Karlheinz Schwarz,
  Techn. Universit\"at Wien, Austria},\ \bibinfo {year} {2001})\BibitemShut
  {NoStop}%
\bibitem [{\citenamefont {Noda}\ \emph {et~al.}(2007)\citenamefont {Noda},
  \citenamefont {Kimura}, \citenamefont {Kamada}, \citenamefont {Ishikawa},
  \citenamefont {Kobayashi}, \citenamefont {Wakabayashi}, \citenamefont {Sawa},
  \citenamefont {Ikeda},\ and\ \citenamefont {Kohn}}]{Noda2007}%
  \BibitemOpen
  \bibfield  {author} {\bibinfo {author} {\bibfnamefont {Y.}~\bibnamefont
  {Noda}}\ \emph {et~al.},\ }\href@noop {} {\bibfield  {journal} {\bibinfo
  {journal} {J. Kor. Phys. Soc.}\ }\textbf {\bibinfo
  {volume} {51}},\ \bibinfo {pages} {828} (\bibinfo {year} {2007})}\BibitemShut
  {NoStop}%
\bibitem [{\citenamefont {Kagomiya}\ \emph {et~al.}(2003)\citenamefont
  {Kagomiya}, \citenamefont {Matsumoto}, \citenamefont {Kohn}, \citenamefont
  {Fukuda}, \citenamefont {Shoubu}, \citenamefont {Kimura}, \citenamefont
  {Noda},\ and\ \citenamefont {Ikeda}}]{Kagomiya2003}%
  \BibitemOpen
  \bibfield  {author} {\bibinfo {author} {\bibfnamefont {I.}~\bibnamefont
  {Kagomiya}}\ \emph {et~al.},\ }\href {\doibase 10.1080/00150190390206347}
  {\bibfield  {journal} {\bibinfo  {journal} {Ferroel.}\ }\textbf
  {\bibinfo {volume} {286}},\ \bibinfo {pages} {167} (\bibinfo {year}
  {2003})}\BibitemShut {NoStop}%
\end{thebibliography}
%References/RXS,References/crystallography

%merlin.mbs apsrev4-1.bst 2010-07-25 4.21a (PWD, AO, DPC) hacked
%Control: key (0)
%Control: author (8) initials jnrlst
%Control: editor formatted (1) identically to author
%Control: production of article title (-1) disabled
%Control: page (0) single
%Control: year (1) truncated
%Control: production of eprint (0) enabled
%

\end{document}